\documentstyle[11pt,newpasp,twoside,epsf]{article} 
\def\beq{\begin{equation}}
\def\eeq{\end{equation}}
\def\beqa{\begin{eqnarray}}
\def\eeqa{\end{eqnarray}}
\def\edcomment#1{\iffalse\marginpar{\raggedright\sl#1\/}\else\relax\fi} 
\reversemarginpar 

\begin{document} 
\title{Cosmology and the Stellar Halo}
\author{James S. Bullock\footnote{}}
\affil{Harvard-Smithsonian Center for Astrophysics, 60 Garden St. MS 51, 
Cambridge,  MA 02138, USA; jbullock@cfa.harvard.edu \\ $^1$ Hubble Fellow}

\author{Kathryn V. Johnston} 
\affil{Van Vleck Observatory, Wesleyan University, Middletown, 
CT 06459-0123, USA; kvj@astro.wesleyan.edu} 

\begin{abstract} 
If  the favored hierarchical cosmological  model is  correct, then the
Milky  Way system  should    have accreted and  subsequently   tidally
destroyed $\sim  100$ low-mass galaxies in  the past $\sim 12$Gyr.  We
model  this process using a hybrid  semi-analytic plus N-body approach
and show that  the disrupted systems  lead naturally  to stellar halos
with masses and density profiles much like the stellar halo of our own
Galaxy. We present predictions for the properties of stellar halos and
show that ours is likely dominated by substructure beyond $\sim 50$kpc
and  more spatially smooth  within that  radius.  The average  stellar
halo density profile is expected to drop off  with radius more quickly
than that of the  dark matter because the stellar  halo is formed from
the  most tightly bound material  in infalling systems,  while the
majority of
the accreted dark matter is  stripped and deposited at larger  radii.
We argue that  stars associated with the inner   halo should be  quite
different  chemically from stars    in surviving satellites and   also
different from  stars in the  outer halo or  those liberated in recent
disruption events.   We  discuss how deep  halo   surveys and chemical
probes may be useful tools for uncovering evidence of accretion.
Searches of  this  kind offer a direct  test  of whether cosmology  is
indeed hierarchical on small scales.
\end{abstract}

\section{Introduction}

The  LCDM model  of  structure formation  predicts that   the dark
matter   halos of  galaxies like the   Milky  Way  were formed by  the
continual accretion of lower-mass systems.  While some fraction of the
accreted dark  halos are expected to survive  as subhalos, most of the
extended dark  halo is built  up from merging  systems that 
are tidally  destroyed (Zentner \&  Bullock
2003).  Similarly, if  the  accreted systems   contain dissipationless
stars,  e.g.  in the form  of a dwarf  galaxy, then this process leads
naturally   to the formation of an    extended stellar halo.  The LCDM
expectation  for the formation of the  stellar  halo is reminiscent of
the Searle \& Zinn  (1978) scenario of  chaotic accretion, and has the
positive  feature that the expected accretion  history  is well-known
theoretically,  at   least  for  the  dark  matter.  In   this
contribution we describe efforts to model the formation of the stellar
halo  in the  context LCDM cosmology using   a  set  of idealized  N-body
simulations.  We  make predictions for  the abundance  of stellar halo
substructure as a function of radius and argue that deep probes of the
stellar  halo  offer  a  unique test   of the   hierarchical nature of
structure formation on small scales.

Modeling the stellar halo from a cosmological context is non-trivial.
Fully self-consistent cosmological simulations
allow the  dark matter dynamics   to be followed   in detail, but even
simulations  of millions of particles   will only have  a few thousand
particles in the very largest satellites  (e.g.  Helmi, White, \& Springel 2002).
Moreover,  such  simulations  are computationally intensive  and hence
prohibit  examining  more   than a   handful  of  halos.  Analytic  or
semi-analytic approaches (e.g.  Bullock  et al. 2001, Johnston, Sackett
\&  Bullock  2001) allow   the  production of   many  halos,  but only
approximate the dynamics.

In this study we  take a hybrid  approach, and use accretion histories
generated semi-analytically within a  cosmological model to inform  us
of the luminosity, mass and accretion times of  satellites, as well as
the instantaneous mass of the parent  galaxy. An N-body simulation for
each accreted satellite is then  run within an  analytic model for the
parent.   This approach  allows  a  vast  decrease in  the  necessary
computing time,  while  at the  same  time allows us to follow the dynamics in
detail.  Our approach  is limited in  its ability to  correctly follow
major  mergers, so our  model results are most  accurate  for the outer
halo ($r \ga 20$kpc).   The  inner halo  formed  early on ($t  \ga
8$Gyr ago) during a period of rapid  mass accretion when major mergers
were  common.  In addition,  the detailed phase-space structure of the
inner  halo was  likely influenced  by  the  formation process  of the
Galaxy itself, which is rather  poorly understood.  For this reason we
focus mostly on the outer  halo, which should   have formed from  more
recently-accreted systems  ($t \la   8$Gyr).  Fortunately,  the vast
majority of expected  stellar substructure comes from accretion events
that occurred late, after the last $> 10\%$ merger event.

We describe the method more fully in \S2 and  our results and
conclusions in \S3.
In what  follows we assume an  LCDM cosmology
with   $\Omega_m = 1 -  \Omega_{\Lambda}  =  0.3$, $h=0.7$, $n=1$, and
$\sigma_8 = 0.9$.  A more thorough presentation of this work will
be given elsewhere (Johnston \& Bullock; Bullock \& Johnston, in
preparation).

\section{Methods}

\subsection{Cosmological Context and Semi-Analytic Star Formation}

We set the mass accretion and subhalo  accretion histories of our host
halos using Extended-Press-Schechter (e.g. Lacey \& Cole
1993) and  employ the merger tree  algorithm proposed by Somerville \&
Kolatt (1999).  This allows us to  generate  mass accretion
histories of our host halos as a function of
time, as shown in Figure 1 for four example realizations
(upper solid  lines).  We demand that the hosts at $z=0$ 
have a mass of $1.4 \times 10^{12} M_{\odot}$.  The merger
tree also  provides a list of  the masses, $M_{\rm  s}$, and accretion
redshifts, $z_{\rm  m}$, of all  subhalos that merge  to form the host
halo.

Assigning luminosities and stellar masses to each accreted dark matter
halo is a fundamental step in any attempt to model the stellar halo in this
context.  We do so by modeling star formation in each system 
based on an estimate for its total mass in cold gas:
 $\dot{M_*} = M_{\rm gas}/t_*$, where $t_*$ is
the star formation  timescale.  For massive  systems, the mass in cold
gas accreted in a  given time is proportional  to the dark matter mass
accreted:  $M_{\rm gas}   =  f_b  (1-f_{\rm hot})  M_{\rm  dm}$.
Here $f_b=0.13$ is the baryon fraction, and $f_{\rm hot} < 1$ parameterizes
the  amount of hot gas  that might arise  from inefficient cooling and
feedback.   For low-mass systems ($<   30 {\rm km}$s$^{-1}$), 
we  solve the  dwarf
satellite problem (e.g.  Klypin et al.  1999) following the suggestion
of Bullock, Kravtsov \&  Weinberg (2000), and  assume that  the total
amount of gas available  to form stars  is proportional to the mass in
place at the epoch of reionization $z_{re} = 10$.  Remarkably, we find
that  the choices $t_* =  15$ Gyr and  $f_{\rm hot}=0.85$ allow us to
roughly reproduce the main  characteristics  of our Galaxy,  including
the stellar halo, disk, and bulge masses, as well as the observed
star formation rates and gas mass fractions of isolated
dwarf galaxies and Local Group irregulars (see Johnston \& Bullock, 
in preparation).

Crucial in this agreement is the rather long star formation timescale
we have assumed for low-mass systems, $t_* = 15$Gyr.  This means that
systems that are accreted early tend to have high gas mass fractions,
and  contribute more to the   main galaxy than   the to stellar halo.
Indeed, the observed  ongoing star formation  in many  dwarf galaxies
effectively demands that the star formation  timescales must be long,
at least in an average sense (e.g. Grebel 2001). 
 The  star formation timescale is
likely shorter  in larger-mass  systems  like  the Milky   Way  host,
perhaps because   the gas can  reach   higher densities  and feedback
effects are less important in stabilizing the star formation.  Figure
1 shows the expected evolution of the disk,  bulge, and stellar halo
components for four  example accretion histories (dotted, dashed, and
thick-dashed from top  to bottom in each  panel).  We have associated
mass accreted in hot gas with disk material, cold gas associated with
disrupted satellites as  bulge material, and  unbound stellar mass as
stellar halo material.   While the disk mass and bulge mass build-up
presented in this figure is somewhat dependent on our assumptions
about gas infall, the general feasibility of our picture for how
galaxies like the Milky Way were assembled is clear.   We go on
to use this simple semi-analytic model to generate the input 
conditions for our more detailed numerical modeling.

\begin{figure}
\plotone{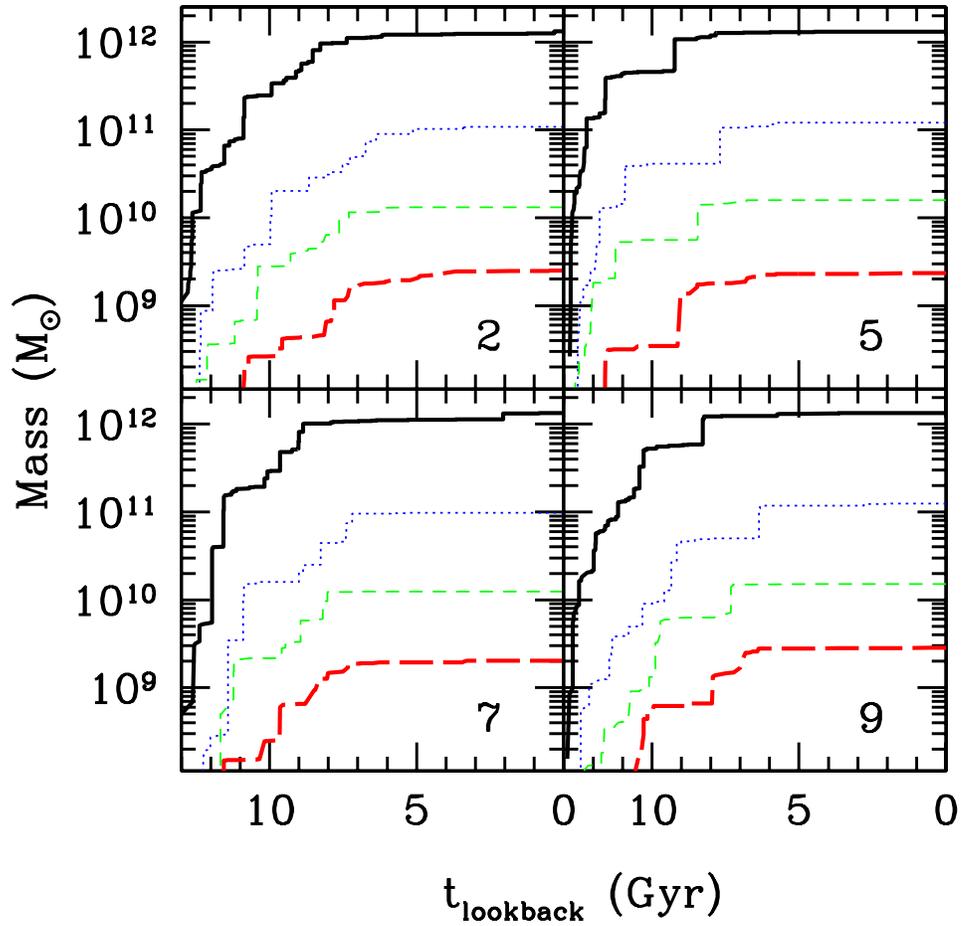} 
\caption{Mass accretion histories for four example 
realizations.  Upper solid lines in each panel show the mass accretion
histories of the dark matter, and the lowest thick long-dashed lines show
the evolution of the stellar halos.  Estimated disk mass and bulge
mass evolution are shown by dot-dash (second from top) and short dashed lines 
respectively.  }
\end{figure}

\subsection{The parent galaxy}

The  potential of the parent  galaxy is represented  by three analytic
functions, corresponding to bulge, disk, and halo components. The first
two are held at fixed mass and scale. For the halo, we use a spherical
NFW (Navarro, Frenk \& White, 1996) model that grows in mass and scale
according to  the  smoothed accretion  history of the  parent  galaxy,
described above.  We do so following
the prescription developed by  Wechsler et al (2002) to  fit
fully self-consistent N-body simulations.

\subsection{The satellite galaxies}

We find that over course of its history, each Milky-Way-type halo
accretes $\sim 100$ luminous satellites.  We simulate the
orbital evolution of each satellite once accreted
using   10$^5$  particles,  initially
distributed as  an  NFW profile  with mass,  scale and accretion  time
taken from  the  semi-analytically  generated  accretion history.  The
orbits  of the satellites are  chosen  at random so that the
distribution of final orbits 
matches that 
measured by Ghigna  et al (1998)  at the end  point of  a cosmological
N-body simulation. The  self-gravity  of the satellite   is calculated
using a basis-function-expansion code (Hernquist \& Ostriker 1992) and
the   influence of the parent  is  imposed from the analytic functions
described above. In addition, all  particles within two tidal radii of
the satellite feel a drag  force due to dynamical friction, calculated
as suggested by Hashimoto, Funato \& Makino's (2003) simulations.
 
The total mass in luminous matter in  the satellites is assigned based
on the star formation algorithm  described above
(gas mass  is ignored for now.)  Based  on
its  luminosity,  we  assign  each  satellite galaxy  a   King density
profile, with a  core radius set  to match the  distribution of scales
observed  in local dwarf  galaxies (Dekel \&  Woo, 2003).  We assign a
luminosity  to every particle   based on its   energy  at the  time of
accretion.   We do so   using Eddington's formula  (Binney \& Tremaine
1982) for the appropriate King  profile embedded within an equilibrium
NFW halo.  The key element here is that the star  particles tend to be
much more tightly  bound  than than most   of the dark matter   in the
infalling satellite halos.

\section{Preliminary Results}

As shown by the  thick dashed lines  in Figure  1, the  total stellar
halo mass at $z=0$  in our model matches  well what is  expected
for our  Galaxy,  $\sim  2 \times 10^{9} M_{\odot}$,   with  the total  value
varying from galaxy-to-galaxy by a factor of  $\sim 2$.  The top panel
of Figure 2 shows  the resulting stellar  halo profiles derived  from
our  simulations   for four  example halos     (four thick lines,  see
caption).  The thin line, for  reference, shows an $r^{-3}$ power law.
Rough agreement with the expected power-law is evident, although we do
predict a  steep fall-off beyond $\sim  50$kpc.  Some understanding of
this steepening can be gained from the lower panel of Figure 2, which
shows  our derived stellar halo profile  along with the profile of all
of the unbound  dark matter mass in our  simulations.  We see that the
stellar profile falls off more quickly at large radii because the star
particles  tended    to sample the most    tightly-bound  cores of the
infalling halos.  It is those cores that tend to survive deep into the
halo core before becoming disrupted.

\begin{figure}
\plotone{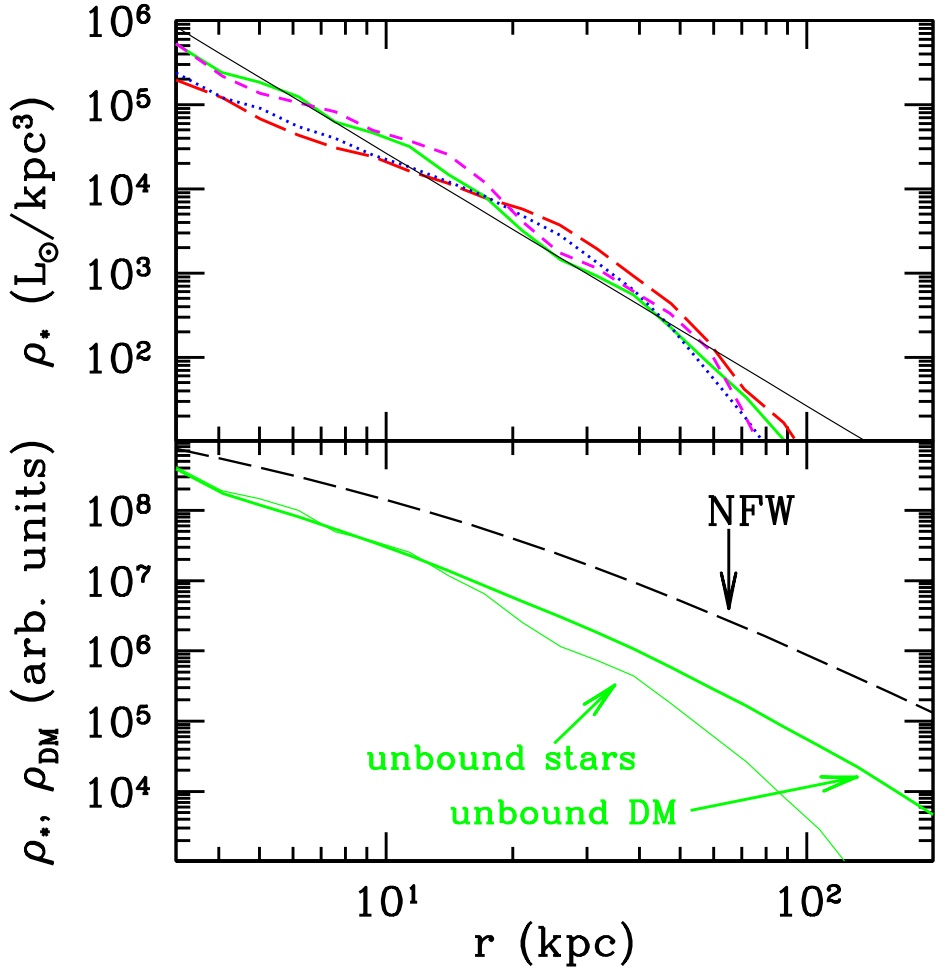}
\caption{Halo profiles.  The upper panel shows the spherically-averaged
stellar halo luminosity profiles for the same four
example realizations shown in Figure 1.  An $r^{-3}$ power-law
is shown by  the thin solid
line.  The bottom panel shows the stellar halo profile for one of our example
halos (thin solid line) compared to the unbound dark matter profile
from our simulations (thick solid line, see arrows).  The normalization
here is arbitrary.  Clearly, the stellar halo falls off more
rapidly at large radius than does the dark matter. 
For comparison, the dashed line shows
the NFW profile we have assumed for the background potential
(shifted up for clarity) at $z=0$.  Interestingly the NFW
profile expected based on full cosmological simulations
is remarkably similar to
that of the cumulative unbound dark matter actually produced
by our idealized simulation, even though this was not forced 
by our method.}
\end{figure}

\begin{figure}
\plotone{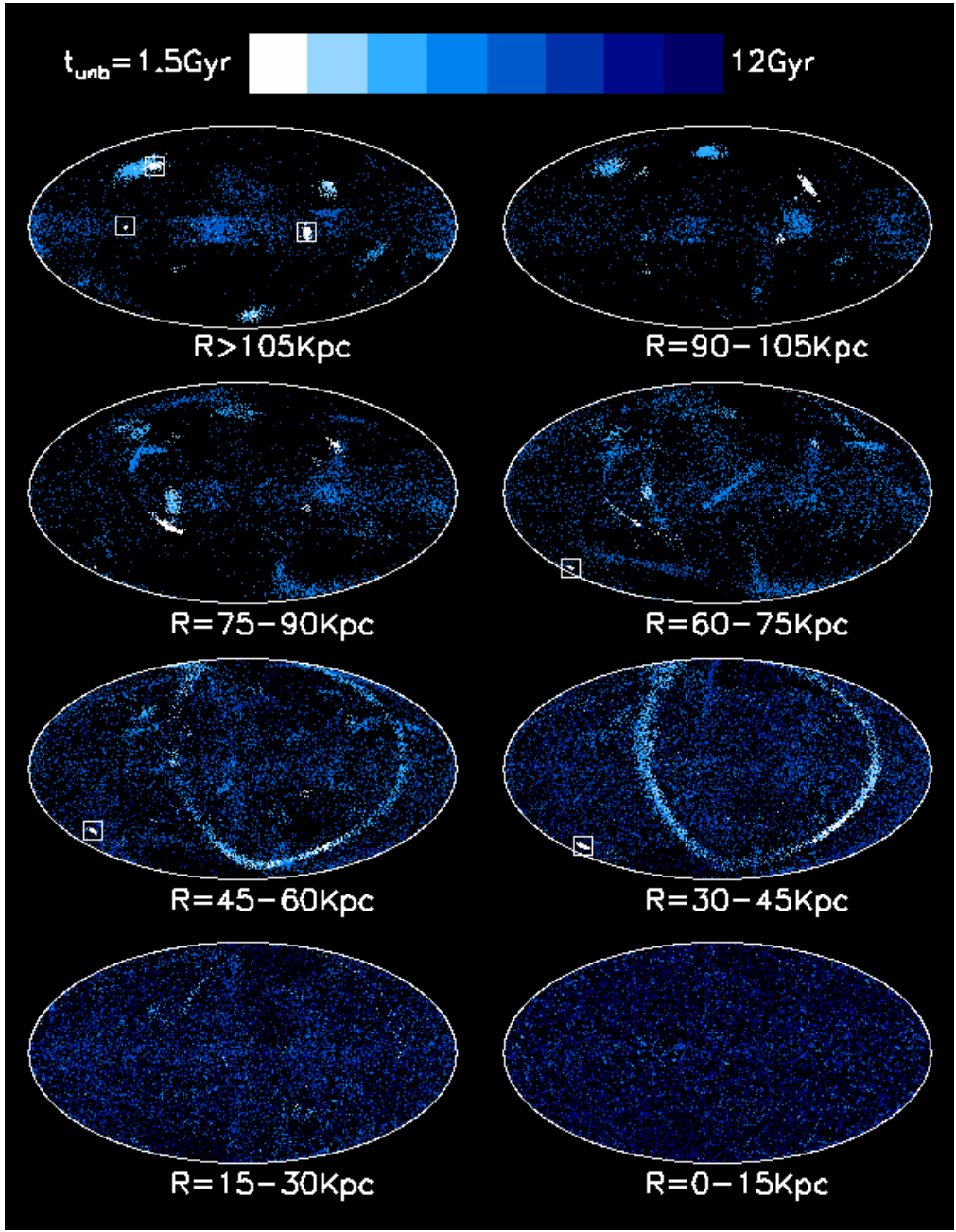}
\caption{One of our model stellar halos shown in all-sky projection within
bins of $15$kpc stepping out in galacto-centric radius.  Each point
represents a tracer star of one per $500 L_{\odot}$ in total V-band
luminosity.  The color-code (shown at the top)
represents the time the particle became unbound from its original
satellite.  Bound particles and recent disruptions are white.  The
boxes signify bound systems at $z=0$.  Substructure is apparent in
this color space, which may provide a reasonable tracer of how chemically
evolved the stars should be.  }
\end{figure}

\begin{figure}
\plotone{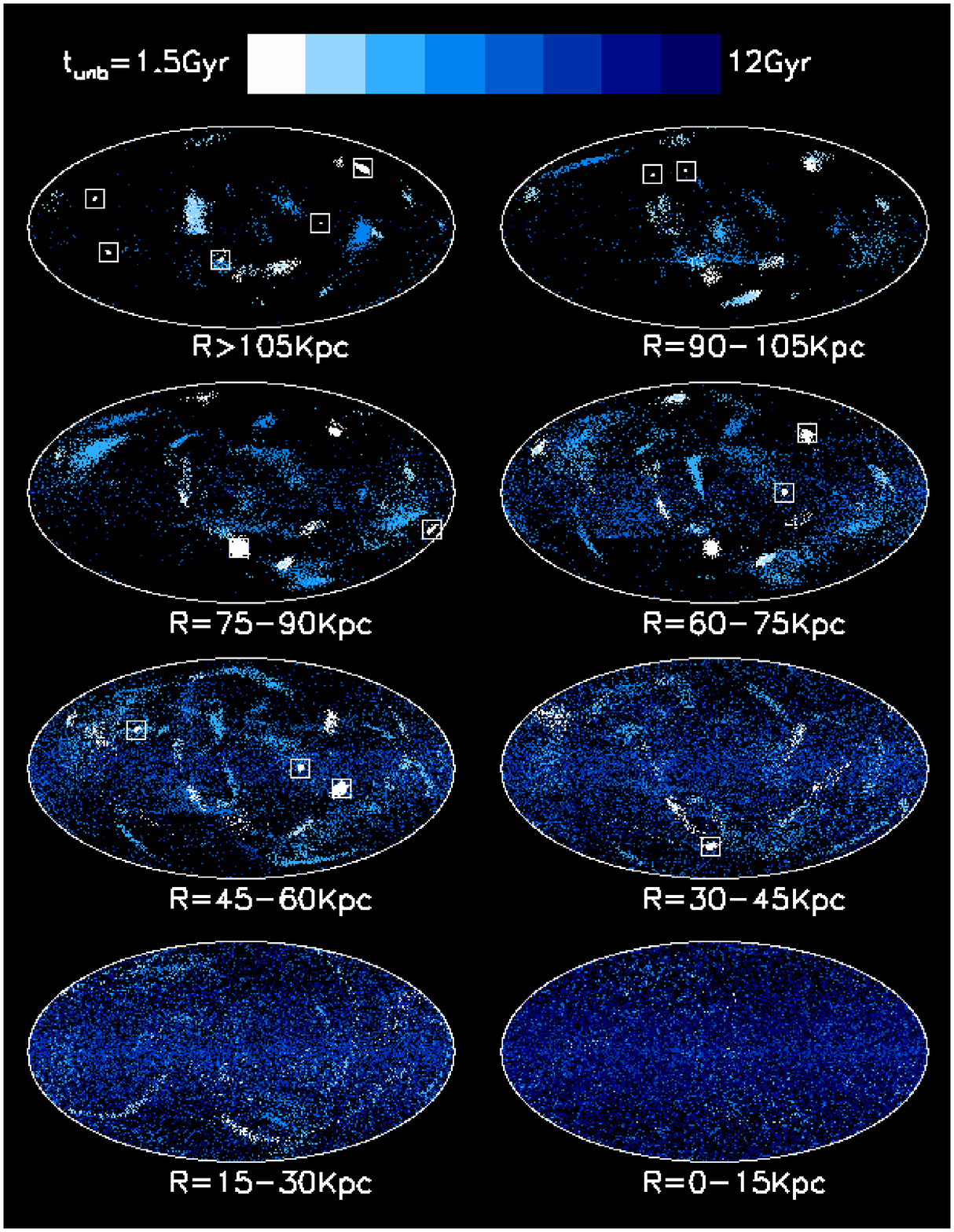}
\caption{A second halo realization shown in all-sky projection.
See the caption of Figure 3 for an explanation.}
\end{figure}

Figures 3 and 4  show  Aitoff sky  projections for two
 of  our stellar halo
realizations.   Projected  in radial  bins,  each  point represents  a
tracer star  that samples 1 in  500 $L_{\odot}$, and we  have assumed a
stellar mass-to-light ratio of 2 in the V-band.  The outer halos, beyond
$\sim  50$kpc, show pronounced  substructure, while  the inner  halos are
more smooth.   We expect  structures of this  kind to be  uncovered by
deep halo surveys like the Deep Giant Star Survey,
Spaghetti Survey, and the Sloan Digital Sky Survey 
(Majewski et al. 2000, Morrison et al. 2000, Yanny et
al. 2000).  We have colored each point based on the time it became
unbound (scale bar at the top),  
and the  substructure is  even  more evident  in this representation.
 Specifically,  more recent disruption events  tend to be seen
as coherent  substructure.   This   is   dramatically apparent  in  the
great-circle stream in Figure 3 
at $r \sim 30-60$kpc, but smaller coherent streams
can be  seen as far  in as $15$kpc, especially in Figure 4.   
If star formation is  ongoing in
satellites after they are accreted, then more chemically evolved stars
will be associated with more recently unbound particles.  This suggests that
chemical probes  may prove useful in substructure  searches.  While not
shown, similar trends are seen even  if we tag particles simply by the
time when the satellite was accreted.  

Note that  the  surviving satellites (boxes) are   likely to  have had
significantly  different chemical evolution  histories than those that
disrupted  early and  contributed  to  the central  halo.   The median
accretion  time for surviving  satellites is $\sim  5$Gyr in the past,
while most  of the stellar halo  was  {\textit{ in place}}  $\ga 8$Gyr
ago.  That the surviving systems are expected to represent a different
population  of dwarfs than those that  get  destroyed may help explain
why most satellite dwarf  abundance patterns do  not look like that of
the average inner stellar halo of the Milky Way (e.g., Shetrone et al.
2001).  Interestingly, studies indicate  that the outer halo may  have
abundances  similar  to those in some  dwarfs  (e.g., Fulbright 2002).
This  trend   is roughly   consistent   with  our   expectations,   as
demonstrated in Figure 3 and especially in Figure 4.

\acknowledgements
JSB  is supported by NASA through  Hubble Fellowship grant
HF-01146.01-A  from the  Space Telescope  Science Institute,  which is
operated by the Association of Universities for Research in Astronomy,
Incorporated, under NASA contract NAS5-26555.
 KVJ was supported  by NASA LTSA grant NAG5-9064, NSF
CAREER   award  AST-0133617   and   a  travel   award  from   Wesleyan
University.  We thank Francisco Prada, David Mart{\'i}nez-Delgado, and
the rest of the organizers for an enjoyable and instructive meeting.

\end{document}